**Postischemic treatment with the cyclooxygenase-2 inhibitor nimesulide reduces blood-brain barrier disruption and leukocyte infiltration following transient focal cerebral ischemia in rats**

Authors: Eduardo Candelario-Jalil [*,‡,1], Armando González-Falcón [‡], Michel García-Cabrera [‡], Olga Sonia León [‡] and Bernd L. Fiebich [*,§]

[*] *Neurochemistry Research Group, Department of Psychiatry, University of Freiburg Medical School, Hauptstrasse 5, D-79104 Freiburg, Germany*

[‡] *Department of Pharmacology, University of Havana (CIEB-IFAL), Havana City 10600, Cuba*

[§] *VivaCell Biotechnology GmbH, Ferdinand-Porsche-Strasse 5, D-79211 Denzlingen, Germany*

Corresponding author:

Dr. Bernd L. Fiebich
Neurochemistry Research Group,
Department of Psychiatry,
University of Freiburg Medical School
Hauptstrasse 5,
Freiburg, D-79104, Germany
Tel.: +49-761-270-6898
Fax: +49-761-270-6916
E-mail: bernd.fiebich@klinikum.uni-freiburg.de

[1] Present address: Department of Neurology, University of New Mexico Health Sciences Center, Albuquerque, NM 87131, USA. E-mail: ECandelario-Jalil@salud.unm.edu

*Abbreviations used:* COX, cyclooxygenase; $PGE_2$, prostaglandin $E_2$; MCAO, middle cerebral artery occlusion; MPO, myeloperoxidase; EB, Evans Blue; VAS, Valeryl Salicylate; BBB, blood-brain barrier; PMN, polymorphonuclear leukocytes; ROS, reactive oxygen species; TTC, 2,3,5-triphenyltetrazolium chloride; MCA, middle cerebral artery.

**Abstract**

Several studies suggest that cyclooxygenase (COX)-2 plays a pivotal role in the progression of ischemic brain damage. In the present study, we investigated the effects of selective inhibition of COX-2 with nimesulide (12 mg/kg) and selective inhibition of COX-1 with valeryl salicylate (VAS, 12-120 mg/kg) on prostaglandin $E_2$ ($PGE_2$) levels, myeloperoxidase (MPO) activity, Evans Blue (EB) extravasation and infarct volume in a standardized model of transient focal cerebral ischemia in the rat. Postischemic treatment with nimesulide markedly reduced the increase in $PGE_2$ levels in the ischemic cerebral cortex 24 h after stroke and diminished infarct size by 48 % with respect to vehicle-treated animals after 3 days of reperfusion. Furthermore, nimesulide significantly attenuated the blood-brain barrier (BBB) damage and leukocyte infiltration (as measured by EB leakage and MPO activity, respectively) seen at 48 h after the initial ischemic episode. These studies provide the first experimental evidence that COX-2 inhibition with nimesulide is able to limit BBB disruption and leukocyte infiltration following transient focal cerebral ischemia. Neuroprotection afforded by nimesulide is observed even when the treatment is delayed until 6 h after the onset of ischemia, confirming a wide therapeutic window of COX-2 inhibitors in experimental stroke. On the other hand, selective inhibition of COX-1 with VAS had no significant effect on the evaluated parameters. These data suggest that COX-2 activity, but not COX-1 activity, contributes to the progression of focal ischemic brain injury, and that the beneficial effects observed with non-selective COX inhibitors are probably associated to COX-2 rather than to COX-1 inhibition.

*Key words:* Cyclooxygenase; prostaglandin $E_2$; stroke; blood-brain barrier; leukocyte infiltration; cerebral ischemia; vasogenic edema; cerebral infarct; neuroprotection

*Running title:* COX-2 is involved in BBB damage after ischemic stroke

**Introduction**

Ischemic stroke disrupts the quality of patients' life, extracts an enormous emotional and physical strain on caregivers, and cost society billions of dollars every year (Taylor *et al.* 1996). Significant progress has been made in dissecting the molecular pathways of excitotoxicity, oxidative stress, apoptosis and neuroinflammation in ischemic neuronal cell death. However, translation of these preclinical results into clinically effective stroke treatments remains a major challenge for the stroke community.

Although a significant amount of ischemic tissue dies in the core of the infarct within few hours after the vessel occlusion, there is evidence showing that the damage in the surrounding tissue (ischemic penumbra) progresses over a relative long period of time (Iadecola and Ross 1997;Dirnagl *et al.* 1999). Thus, pharmacological strategies limiting the delayed phase of the damage are probably more important in stroke therapy, since most of the patients arrive in the emergency room too late for preventing or minimizing the initial damage. Inflammation is one of the mechanisms known to participate in the progression of brain injury (Dirnagl *et al.* 1999;Dirnagl 2004). It has been shown that after several hours of the onset of ischemia, there is a significant disruption of the blood-brain barrier (BBB) followed by a massive infiltration of polymorphonuclear (PMN) leukocytes (Rosenberg *et al.* 1998;Batteur-Parmentier *et al.* 2000;Martin *et al.* 2006). This results in brain edema and microglial activation, and the production of large amounts of pro-inflammatory cytokines, ROS, nitric oxide, among other mediators of neuroinflammation, which exacerbate tissue damage. All these neuroinflammatory mechanisms have been demonstrated to contribute to ischemic brain injury (Barone and Feuerstein 1999;Dirnagl *et al.* 1999;Stanimirovic and Satoh 2000). A large number of studies indicates that blockade of the neuroinflammatory process dramatically reduces ischemic brain injury (Nogawa *et al.* 1997;Nogawa *et al.* 1998;Nagayama *et al.* 1999;Batteur-Parmentier *et al.* 2000;Candelario-Jalil *et al.* 2004;Candelario-Jalil *et al.* 2005;Ikeda-Matsuo *et al.* 2006;Kawano *et al.* 2006).

Two different isoforms of the cyclooxygenase (COX) enzyme, COX-1 and COX-2, have been identified (Smith *et al.* 2000). In addition, a COX-1 splice variant, termed COX-3, has been recently cloned and characterized (Chandrasekharan *et al.* 2002;Snipes *et al.* 2005). Large amounts of free arachidonic acid are released during ischemic brain damage through the concert action of phospholipases (Phillis and O'Regan 2003;Phillis and O'Regan 2004;Muralikrishna and Hatcher 2006). COX-2 inhibition is an attractive pharmacological target since the metabolism of arachidonic acid through the COX pathway produces huge amounts of pro-inflammatory prostanoids and ROS, which are key mediators of the inflammatory process (Smith *et al.* 2000). Numerous studies have found a dramatic increase in COX-2 expression following ischemia (Planas *et al.* 1995;Collaco-Moraes *et al.* 1996;Nogawa *et al.* 1997;Miettinen *et al.* 1997;Sasaki *et al.* 2004), and other insults resulting in neurodegeneration (Hewett *et al.* 2000;Scali *et al.* 2000;Strauss *et al.* 2000;Salzberg-Brenhouse *et al.* 2003;Kawaguchi *et al.* 2005;Hewett *et al.* 2006). However, it is worth noting that COX-2 is linked to synaptic activity, and several healthy neuronal populations express COX-2 under normal conditions (Yamagata *et al.* 1993;Adams *et al.* 1996). Furthermore, COX-2 is rapidly induced after a mild episode of focal ischemia, which does *not* result in neuronal damage (Planas *et al.* 1999).

Several studies have demonstrated that selective COX-2 inhibition or COX-2 gene deletion confers neuroprotection in models of ischemic brain injury (Nogawa *et al.* 1997;Nogawa *et al.* 1998;Iadecola *et al.* 2001a;Sugimoto and Iadecola 2003;Candelario-Jalil *et al.* 2004;Sasaki *et al.* 2004;Candelario-Jalil *et al.* 2005). However, there is debate on the specific role of COX-1 in cerebral ischemia. Some studies have found beneficial effects (Lin *et al.* 2002), others claim that COX-1 is detrimental (Iadecola *et al.* 2001b), while in another report COX-1 gene deletion has been shown not to affect ischemic brain injury (Cheung *et al.* 2002).

In an earlier study, we assessed the relative contribution of each COX isoform to global ischemic brain injury. Interestingly, we found that either inhibition of COX-1 with valeryl salicylate (VAS) or selective inhibition of COX-2 with rofecoxib (Vioxx), potently reduced ischemia-induced neuronal cell death and oxidative stress in the hippocampus (Candelario-Jalil *et al.* 2003b), thus challenging the traditional belief that only COX-2 is involved in neuroinflammation during brain ischemia.

Since there are important differences between global and focal cerebral ischemia in terms of pathophysiological mechanisms involved in tissue damage, and considering that there are no previous reports evaluating the effects of selective blockade of each COX isozyme in a focal cerebral ischemia model in relation to BBB damage, $PGE_2$ accumulation, leukocyte infiltration and vasogenic edema, we decided to conduct the present investigation to study the specific role of each COX isoform in the ischemic brain, using a clinically-relevant model of stroke.

Present findings support the notion that treatment with nimesulide reduces ischemic brain injury, and suggest for the first time that *postischemic* treatment with a COX-2 inhibitor confers a significant protection against the late opening of the BBB, which facilitates PMN leukocyte infiltration into the ischemic brain, and vasogenic edema. Selective inhibition of COX-1 with valeryl salicylate had no significant effect on the evaluated parameters. Together, these data suggest the key role of COX-2, rather than COX-1, in the late progression of tissue damage in *focal* cerebral ischemia.

## Materials and Methods

### Animals

All the experimental procedures were performed strictly according to the regulations of the Havana University's animal ethical committee and the guidelines of the National Institutes of Health (Bethesda, MD, USA) for the care and use of laboratory animals for experimental procedures. Our institutional animal care and use committee approved the experimental protocol. Appropriate measures were taken to minimize pain and distress of animals used in this study. A total of 349 male Sprague-Dawley rats (CENPALAB, Havana, Cuba) weighing 270-320 g at the time of surgery were used in the present study. The animals were quarantined for at least 7 days before the experiment. Animals were housed on bagasse bedding in groups of 2-4 in polycarbonate cages in a room whose environment was maintained at 21-22ºC, 45-50 % humidity and 12 h light/dark cycle. They had free access to rodent pellet chow and water.

### Surgical preparation and procedure for inducing transient focal cerebral ischemia

Rats were anesthetized with chloral hydrate (300 mg/kg body weight, i.p.). Once surgical levels of anesthesia were attained (assessed by absence of hind leg withdrawal to pinch), ischemia was induced by using an occluding intraluminal suture as described before (Koizumi *et al.* 1986;Reglodi *et al.* 2000;Candelario-Jalil *et al.* 2004;Candelario-Jalil *et al.* 2005). Briefly, under microscopic magnification, the anterior neck was opened with a midline vertical incision, and the underlying submandibular gland dissected. Dissection medial to the right sternocleidomastoid muscle exposed the common carotid artery (CCA), and allowed separation of the CCA from the vagus nerve. Both, the CCA and the external carotid artery (ECA) were ligated with a 3-0 silk suture. The pterygopalatine branch of the internal carotid artery was clipped to prevent incorrect insertion of the occluder filament. Arteriotomy was performed in the CCA approximately 3 mm proximal to the bifurcation and a 3-0 monofilament nylon suture (Shenzhen Runch Industrial Corp., China), whose tip had been rounded by being heated near a flame, was introduced into the internal carotid artery (ICA) until a mild resistance was felt (Candelario-Jalil *et al.* 2004;Candelario-Jalil *et al.* 2005). Mild resistance to this advancement indicated that the intraluminal occluder had entered the anterior cerebral artery (ACA) and occluded the origin of the ACA, the middle cerebral artery (MCA) and posterior communicating arteries. The occluding suture was kept in place for 1 h. At the end of the ischemic period, the suture was gently retracted to allow reperfusion of the ischemic region. The incision was closed and animals were allowed to recover from anesthesia and to eat and drink freely. Rectal temperature was maintained at 37 ± 0.5 °C with a heat lamp and electrically heated mat during surgery, stroke, and reperfusion. By using this standardized procedure, we obtained large and reproducible infarcted regions involving the temporoparietal cortex and the laterocaudal part of the caudate putamen in ischemic animals (Candelario-Jalil *et al.* 2004).

**Neurobehavioral testing**

An independent observer blinded to the animal treatment performed the neurological evaluations prior to the sacrifice of the animals according to a six-point scale: 0= no neurological deficits, 1= failure to extend left forepaw fully, 2= circling to the left, 3= falling to left, 4= no spontaneous walking with a depressed level of consciousness, 5= death (Longa *et al.* 1989;Minematsu *et al.* 1992).

**Infarct volume assessment**

The method for quantification of infarct volume was performed exactly as previously reported (Yang *et al.* 1998;Gonzalez-Falcon *et al.* 2003;Candelario-Jalil *et al.* 2004;Candelario-Jalil *et al.* 2005). Briefly, the animals were sacrificed under deep anesthesia and brains were removed, frozen and coronally sectioned into six 2-mm-thick slices (from rostral to caudal, first to sixth) using a rat brain matrix (World Precision Instruments, Sarasota, FL, USA). The brain slices were incubated for 30 min in a 2% solution of 2,3,5-triphenyltetrazolium chloride (TTC) (Sigma Chemical Co., Saint Louis, MO, USA) at 37 °C and fixed by immersion in a 4% paraformaldehyde solution in phosphate-buffered saline pH 7.4. Six TTC-stained brain sections per animal were placed directly on the scanning screen of a color flatbed scanner (Hewlett Packard, HP Scanjet 5370 C) within 7 days. Following image acquisition, the images were analyzed blindly using a commercial image processing software program (Photoshop, version 7.0, Adobe Systems; Mountain View, CA, USA). An investigator blinded to the animal treatment performed the measurements of infarct volume by manually outlining the margins of infarcted areas. The unstained area of the fixed brain section was defined as infarcted. Cortical and subcortical uncorrected infarcted areas and total hemispheric areas were calculated separately for each coronal slice. Total, cortical and subcortical uncorrected infarct volumes were calculated by multiplying the infarcted area by the slice thickness and summing the volume of the six slices. A corrected infarct volume was calculated to compensate for the effect of brain edema. An edema index was calculated by dividing the total volume of the hemisphere ipsilateral to MCAO by the total volume of the contralateral hemisphere. The actual infarct volume adjusted for edema was calculated by dividing the infarct volume by the edema index (Yang *et al.* 1998;Reglodi *et al.* 2000;Yang *et al.* 2000;Candelario-Jalil *et al.* 2004).

**Prostaglandin $E_2$ ($PGE_2$) Enzyme Immunoassay**

Tissue concentration of $PGE_2$, one of the major cyclooxygenase reaction products in the brain (Nogawa *et al.* 1997), was determined using a commercial enzyme immunoassay kit (RPN 222, Amersham Pharmacia Biotech Inc., Piscataway, NJ, USA) according to the instructions of the manufacturer. Unanesthetized animals were decapitated and brains were quickly removed from the skull, and frozen in liquid nitrogen. A 4-mm-thick coronal slice was cut at the level of the optic chiasm, and the cerebral cortex and striatum from both hemispheres were quickly dissected out on a chilled plate (placed on powdered dry ice). The tissue was homogenized in ice-cold 50 mM Tris-HCl (pH 7.4), and extracted with 100 % methanol (Powell 1982). After centrifugation, the supernatant was diluted with acidified 0.1 M phosphate buffer (pH 4.0; final methanol concentration, 15%) and applied to activated octadecylsilyl (ODS)-silica reverse-phase columns (Sep-Pak C18, Waters Associates, Milford, MA, USA). In order to improve recovery, C18 cartridges were rinsed, in this order, with water-acetonitrile-chloroform-acetonitrile-water. After that, cartridges were activated with methanol, ethanol and water. The columns were rinsed with 5 mL of distilled water followed by 5 mL of n-hexane, and $PGE_2$ was eluted twice with 2 mL of ethyl acetate containing 1% methanol. The ethyl acetate fraction was evaporated and resuspended in 1 mL of the buffer provided with the kit. $PGE_2$ concentration was determined spectrophotometrically after incubation with tracer and $PGE_2$ monoclonal antibody in a 96-well plate following the manufacturer's instructions.

**Evaluation of blood-brain barrier (BBB) integrity**

The integrity of the BBB was investigated using Evans blue (EB) dye as a marker of albumin extravasation as reported previously (Belayev *et al.* 1996;Belayev *et al.* 1998;Asahi *et al.* 2001;Matsuo *et al.* 2001). Evans blue (2% in saline, 4 mL/kg) was injected to rats via the tail vein under diethyl ether anesthesia at different times (time course experiment) or at 46 h after the onset of MCAO (for evaluating the effect of nimesulide and valeryl salicylate; see below Section *'Treatment groups and drug administration'*). Two hours after the EB injection, the rats were anesthetized with chloral hydrate and perfused with physiological saline through the left ventricle until colorless perfusion fluid was obtained from the *vena cava* (150-200 mL of saline). Brain samples from ipsilateral and contralateral hemispheres were dissected out (cerebral cortex, striatum and rest of the hemisphere) for local measurement of EB extravasation. Samples were immediately weighed and placed in cold 50% trichloroacetic acid solution. Following homogenization and centrifugation (10 min at 12000 rpm), the extracted dye was measured spectrofluorimetrically, as described before (Belayev *et al.* 1996;Belayev *et al.* 1998). The quantitative calculation of the dye content in each brain area was based on external standards in the same solvent. The tissue content of EB was quantified from a linear standard curve derived from known amounts of the dye (25-1000 ng of EB/mL), and expressed per gram of wet tissue. This procedure has been widely used to evaluate BBB breakdown following ischemic stroke (Belayev *et al.* 1996;Kondo *et al.* 1997;Belayev *et al.* 1998;Matsuo *et al.* 2001;Ding-Zhou *et al.* 2003).

**Cerebral Tissue Myeloperoxidase Content**

Myeloperoxidase (MPO), a lysosomal enzyme specific to leukocyte granules, has been used as an index of PMN leukocyte accumulation in the ischemic tissue. The method used to quantify MPO activity from rat brain samples was similar to that recently described by others (Batteur-Parmentier *et al.* 2000;Matsuo *et al.* 2001;Couturier *et al.* 2003;Ding-Zhou *et al.* 2003;Martin *et al.* 2006). Briefly, rats were anesthetized with diethyl ether and perfused transcardially with ice-cold physiological saline to flush all blood components from the vasculature. After brain dissection, samples were immediately weighed and quickly frozen in liquid nitrogen. Each sample was homogenized in 20 volumes of 5 mM potassium phosphate buffer (4°C, pH 6.0) followed by centrifugation at 30,000 x *g* for 30 minutes at 4°C. The supernatant was discarded and the pellet was washed again as described above. After decanting the supernatant, the pellet was extracted by suspension in 10 times the volume of 0.5% hexadecyltrimethylammonium bromide (HTAB, Sigma-Aldrich) in 50 mM potassium phosphate buffer (pH 6.0) at 25°C. HTAB is a detergent that releases the MPO enzyme from leukocyte granules. The samples were frozen on dry ice, and 3 freeze/thaw cycles were then performed to further disrupt granules, with sonication between cycles. After the last sonication, samples were incubated at 4°C for 20 min, and centrifuged at 12,500 x $g$ for 15 min at 4°C. MPO activity in the supernatant was assayed as described before (Biagas $et$ $al.$ 1992). Briefly, 100 µL supernatant was mixed with 2.9 mL 50 mM potassium phosphate buffer containing 0.167 mg/mL o-dianisidine dihydrochloride (Sigma) and 0.0005% hydrogen peroxide. The change in absorbance at 460 nm was recorded spectrophotometrically (Pharmacia LKB) at 15-second-intervals for 2 min in triplicate. MPO activity was calculated as the mean of the three readings. One unit (U) of MPO activity was defined as the amount that degraded 1 µmol hydrogen peroxide per minute at 25°C, and was normalized on the basis of wet tissue weight (U/g wet tissue).

**Treatment groups and drug administration**

Nimesulide and VAS were dissolved in a 2% polyvinylpyrrolidone (PVP) solution in saline as reported before (Candelario-Jalil $et$ $al.$ 2002;Candelario-Jalil $et$ $al.$ 2003b;Candelario-Jalil $et$ $al.$ 2003a;Candelario-Jalil $et$ $al.$ 2004). These inhibitors were given intraperitoneally starting either immediately after ischemia or in a 6 h delayed treatment. Additional doses were given at 6, 12, 24, 36 and 48 h after stroke (for evaluating infarct volume and neurological deficits at 3 days). This treatment schedule and dosage were based on the pharmacokinetic profiles of nimesulide (Toutain $et$ $al.$ 2001), and on our previous experience with these COX inhibitors in models of cerebral ischemia, where different doses and treatment regimes were studied (Candelario-Jalil $et$ $al.$ 2002;Candelario-Jalil $et$ $al.$ 2003b;Candelario-Jalil $et$ $al.$ 2004). In the experiments in which animals were sacrificed after 48 h of reperfusion for evaluating BBB damage, leukocyte infiltration, and edema, drugs were given either immediately or starting 6 h after MCAO, with additional doses at 6, 12, 24 and 36 h. In the experiment evaluating the effects of these COX inhibitors on $PGE_2$ formation in the ischemic brain, animals were treated with vehicle, nimesulide (12 mg/kg) or VAS (12 or 120 mg/kg) starting immediately after ischemia, and additional doses were administered at 6, 14 and 22 h after induction of MCAO. Rats were sacrificed for $PGE_2$ analysis 2 h after the last injection. This treatment paradigm is similar to that employed in previous studies (Nogawa *et al.* 1997;Candelario-Jalil *et al.* 2004).

**Statistical analysis**

Data are presented as mean ± S.D. Infarct volume, $PGE_2$ data, MPO activity, EB content and percent of edema were analyzed using t-test (2 groups) or one-way ANOVA with *post-hoc* Student-Newman-Keuls test (multiple comparison). Neurological deficit scores were analyzed by Kruskal-Wallis nonparametric ANOVA followed by the Dunn test (multiple comparison) or Mann-Whitney test for analysis of individual differences. In all statistical tests, differences were considered significant when $p < 0.05$.

**Results**

*Temporal evolution of the ischemic lesion and $PGE_2$ production after transient occlusion of the middle cerebral artery in the rat*

All animals studied had visible ischemic lesions in the MCA territory after 1 h ischemia and 6 h of reperfusion (earliest time point studied). There was a significant increase in cortical infarct volume over the time, as observed in Fig. 1A. The size of the infarct in the cerebral cortex reached its maximal values by 3 days of recirculation, and no statistically significant differences were seen between 3 and 4 days of reperfusion (Fig. 1A). Infarct size in the cortical areas at 3-4 days of reperfusion almost doubled that seen after 24 h in this model of temporary MCA occlusion. Unlike the cerebral cortex, subcortical regions became necrotic within 6 h of reperfusion, and no appreciable increases in the lesion size over time were noticed in this model of ischemic stroke. Total infarct volume, evaluated at several times after the withdrawal of the occluding filament, followed a similar pattern to that observed in the cerebral cortex. As depicted in Fig. 1A, the growth of the ischemic lesion in the cerebral cortex accounted for the overall temporal increase in total infarct volume.

In our next experiments, we investigated the time course of $PGE_2$ production by the ischemic brain. By 12 h of recirculation, $PGE_2$ levels in the ischemic cerebral cortex began to be significantly increased with respect to the contralateral (intact) side. Maximal $PGE_2$ production was observed at 24 h (p = 0.00127), although $PGE_2$ concentrations remained significantly ($p < 0.05$) augmented up to 72 h of reperfusion in the cerebral cortex of animals suffering from ischemic stroke, as shown in Fig. 1B. No significant increases in subcortical $PGE_2$ levels were observed in our study at any recirculation time (data not shown).

*Time course of the BBB damage, edema formation, and leukocyte infiltration following a temporary occlusion of the middle cerebral artery in the rat*

We studied the temporal changes in BBB damage by quantifying the extravasated Evans blue (EB) concentrations in the ischemic brain (cerebral cortex and subcortical areas) after several times of reperfusion following the ischemic episode. A very early significant ($p < 0.05$) BBB breakdown was observed in both cortical and subcortical regions by 2 h of recirculation, as shown in Figs. 2A and 2B. Unlike the cerebral cortex, in the subcortical areas, this initial EB leakage persisted till 6 h, showing a significant increase in EB extravasation as compared to the contralateral side (Fig. 2B). Interestingly, a late opening in the BBB was then demonstrated 24-72 h post-reperfusion in all ischemic regions investigated. The BBB disruption was maximal at 48 h, and this late opening showed a similar pattern in both the cerebral cortex and subcortical areas (Figs. 2A and 2B). Significant formation of brain edema was seen starting after 12 h of reperfusion, and reached maximal values by 48-72 h post-recirculation, as depicted in Fig. 2C.

The infiltration of PMN leukocytes into the brain parenchyma is another important hallmark of the postischemic neuroinflammatory component of cerebral ischemia. By measuring the activity of an enzymatic marker of these inflammatory cells (MPO), we assessed the degree of leukocyte infiltration into the cerebral cortex and subcortical regions following the focal ischemic event. With respect to the contralateral side, a significant increase ($p < 0.05$) in MPO activity was observed as early as 2 h of reperfusion, and persisted until 3 days following the withdrawal of the nylon filament occluding the MCA. Maximal leukocyte infiltration was seen at 24-48 h in the cerebral cortex, and at 12-24 h post-reperfusion in the subcortical areas (Figs 3A and 3B).

No significant changes in EB leakage, edema and MPO activity, were observed between the ipsilateral and contralateral side of rats that underwent a sham operation. Furthermore, no significant alterations in these variables were seen, when comparing the contralateral side of ischemic animals with respect to the contralateral side of sham-operated rats (data not presented).

*Effects of selective inhibition of COX-1 and COX-2 on infarct volume and neurological deficits*

Infarct volume was assessed with the vital TTC staining at 3 days of reperfusion after the ischemic event. As expected, based on findings of lesion size from our previous studies (Candelario-Jalil *et al.* 2004;Candelario-Jalil *et al.* 2005), treatment with the selective COX-2 inhibitor nimesulide significantly (p<0.05) reduced cortical and subcortical infarct size as presented in Table 1. Nimesulide was still effective in limiting ischemic damage when the first treatment began 6 h after ischemia. As far as the total infarct is concerned, nimesulide's effect on total infarction is mainly the result of its potent reduction in the cortical rather than subcortical lesion (Table 1). The COX-1 selective inhibitor valeryl salicylate (VAS) failed to confer any protective effect, when administered immediately after ischemia, or in a delayed fashion. This lack of effect was seen in both cortical and subcortical areas of the infarct (Table 1).

A scattergram of the neurological scores per treatment group is presented in Fig. 4. Nimesulide was able to produce a significant reduction (p<0.05; Mann-Whitney test) in the neurological deficits seen in the animals after ischemia as compared to vehicle-treated rats. This effect was observed even with the 6-h delayed treatment paradigm (Fig. 4). However, the COX-1 inhibitor VAS conferred no protective effect against stroke-induced neurological impairment, as shown in Fig. 4.

*Prostaglandin $E_2$ concentrations in the ischemic cortex are reduced by nimesulide but not by the COX-1 inhibitor VAS*

We investigated the effect of selective inhibition of either COX-1 or COX-2 on $PGE_2$ levels in the ischemic cerebral cortex after 24 h of reperfusion. As compared to the contralateral side or to the sham-operated animals, occlusion of the MCA resulted in a dramatic increase in the COX product $PGE_2$ (Fig. 5). Administration of the COX-2 inhibitor nimesulide produced a significant protective effect against ischemia-induced $PGE_2$ accumulation in the cerebral cortex, keeping $PGE_2$ concentrations at the basal level (compared to ipsilateral side of sham-operated animals). Treatment with a similar dose of VAS (12 mg/kg) failed to prevent $PGE_2$ increase in the ischemic brain, although a slight decrease in $PGE_2$ levels in the ipsilateral cortex was observed when the dose was increased to 120 mg/kg. Both doses of VAS were able to significantly reduce $PGE_2$ concentrations in the contralateral side when compared to the group of animals, which underwent the sham operation (Fig. 5).

*COX-2 inhibition protects against BBB disruption following ischemic stroke*

In our next experiments, the effect of selective inhibition of COX-1 or COX-2 on BBB breakdown was studied in this model of focal cerebral ischemia. We decided to perform these experiments after 48 h of reperfusion, since at this time point EB leakage was maximal (Fig. 2). Nimesulide (12 mg/kg, i.p.), but not VAS, significantly (p<0.01) attenuated EB extravasation in the ischemic cortex even when the first treatment was delayed until 6 h after ischemia (Fig. 6). No protective effect of these COX inhibitors was observed on the ischemia-mediated BBB damage in the subcortical areas (Fig. 6).

*Leukocyte infiltration and vasogenic edema are potently reduced in nimesulide-treated rats*
We were also interested in the effects of nimesulide and VAS on brain leukocyte infiltration and edema associated to the ischemic injury. Selective inhibition of COX-2 with nimesulide conferred a potent protective effect against stroke-induced leukocyte infiltration into the cerebral cortex (Table 2), as evaluated by the MPO activity assay in the ischemic tissue at 48 h after the withdrawal of the nylon filament occluding the MCA. The neuroprotective efficacy of nimesulide was still observed when the first dose was given 6 h after the occlusion of the MCA (Table 2). However, no effect of nimesulide was seen in the subcortical regions of the infarct.

The protection of the BBB observed in the animals given nimesulide (Fig. 6) translated into a significant reduction in the vasogenic edema (p<0.01). As presented in Table 2, nimesulide potently limited the edema formation at 48 h after ischemia, when administered immediately after MCAO or in a delayed fashion (Table 2). The selective COX-1 inhibitor VAS showed no effect on ischemia-induced leukocyte infiltration and edema (Table 2).

**Discussion**
The availability of selective inhibitors of the COX isozymes provides a powerful pharmacological tool in order to dissect the relative contribution of each isoform to the inflammatory process *in vitro* and *in vivo*. Using this approach, we previously studied the role of each COX isoenzyme in CA1 hippocampal neuronal death in a model of temporary global cerebral ischemia, demonstrating an important role of *both* COX isoforms in ischemia-induced oxidative damage and neurodegeneration (Candelario-Jalil *et al.* 2003b). Interestingly, in the present study, we found that only COX-2 activity is responsible for the evolution of focal cerebral ischemic injury in relation to $PGE_2$ accumulation, BBB disruption, leukocyte infiltration and vasogenic edema, well-known factors involved in brain damage. The different model of cerebral ischemia (global vs. focal) may explain the differences between our two studies. These new observations shed more light into the specific role of the COX/$PGE_2$ pathway in ischemic brain injury, and might have important implications for the potential use of COX inhibitors or agents modulating $PGE_2$ formation/signaling in different clinical settings of cerebral ischemia.

The pharmacological effects of nimesulide have been attributed to its ability to selectively inhibit the COX-2 isoform (Famaey 1997). However, nimesulide is *not* a highly selective COX-2 inhibitor. Thus, we cannot rule out the possibility that some degree of COX-1 inhibition is in play in the ischemic animals treated with nimesulide. However, the dose and administration regime used in the present study failed to significantly reduce basal $PGE_2$ levels in the intact side when compared to sham-operated controls (Fig. 5). Unlike nimesulide, the COX-1 inhibitor VAS significantly reduced basal levels of $PGE_2$ in the cerebral cortex (Fig. 5). If a pharmacologically relevant degree of COX-1 inhibition occurs after nimesulide treatment, one might expect a significant reduction in basal $PGE_2$ levels. These findings suggest that the beneficial effects of nimesulide are due to selective inhibition of COX-2, rather than to a non-selective inhibition of both COX isoforms.

The present study has assessed for the first time the contribution of each COX isoform to $PGE_2$ formation, BBB damage and infiltration of PMN leukocytes in an *in vivo* model of temporary cerebral ischemia. Furthermore, to the best of our knowledge, a detailed time course of $PGE_2$ formation, and its relation to the evolution of brain infarct, had not been previously investigated.

Restoration of cerebral blood flow after ischemia may cause damage to the BBB, exacerbate brain edema, and cause leukocyte infiltration (Chen *et al.* 1995;Batteur-Parmentier *et al.* 2000). Thus, reperfusion injury is a potentially hazardous complication of surgical revascularization, temporary intraoperative cerebrovascular occlusion, or thrombolytic therapy for acute stroke. In the center of the lesion, severe ischemia leads to rapid necrosis (Fig. 1A), but in the surrounding penumbral regions, the tissue damage evolves slowly over many hours/days (Marchal *et al.* 1996). Therapeutic strategies to limit infarct size and improve functional outcome after acute stroke are aimed at rescuing this potentially reversible ischemic region (Fisher 1997). In humans, infarct expansion at the expense of potentially viable tissue has been documented even 24 h after stroke onset (Baird *et al.* 1997).

Post-ischemic inflammation has recently emerged as an important factor responsible for the evolution of the ischemic brain injury. In this regard, present findings indicate that COX-2 selective inhibition with nimesulide blocked late $PGE_2$ production, ischemia-induced BBB breakdown, leukocyte infiltration and edema formation. It is worth noting that this protective effect was observed even when the first treatment was delayed up to 6 h after the onset of MCAO (Tables 1 and 2, Fig. 6). These results, together with the finding that selective COX-1 inhibition with VAS is *not* protective, tempt us to suggest that COX-2 activity plays a role of paramount importance in the progression of focal ischemic brain injury. The lack of effect of VAS in the present study could not be explained by its poor penetrability into the brain, since we proved this inhibitor to significantly reduce basal $PGE_2$ levels in the non-ischemic hemisphere (Fig. 5). In addition, this COX-1 inhibitor has been demonstrated before to exert neuroprotective efficacy in global cerebral ischemia (Candelario-Jalil *et al.* 2003b) at similar or even lower doses than the ones tested in the present investigation.

The wide therapeutic window of protection of COX-2 selective inhibitors has been demonstrated in models of cerebral ischemia (Nogawa *et al.* 1997;Nagayama *et al.* 1999;Candelario-Jalil *et al.* 2002;Candelario-Jalil *et al.* 2003a;Candelario-Jalil *et al.* 2003b;Sugimoto and Iadecola 2003;Candelario-Jalil *et al.* 2004;Sasaki *et al.* 2004) and traumatic brain injury (Gopez *et al.* 2005). The wide therapeutic time window of protection of COX-2 inhibitors in ischemic stroke has very important implications in the clinical practice. One of the most important predictors of clinical success in stroke is time to treatment. Most patients with ischemic stroke reach the hospital several hours after the onset of symptoms, a time at which most therapeutic strategies are no longer effective, or could worsen cerebral injury, as is the case of thrombolysis, which is contraindicated at later times due to increased cerebral hemorrhage (Clark *et al.* 1999;Davis *et al.* 2006).

It has been previously shown that the normal function of the BBB is altered by ischemia (Ballabh *et al.* 2004;Hawkins and Davis 2005). Increase in BBB permeability is associated with severe ischemic damage, occurring with some delay after the initial insult. The biphasic opening of the BBB observed in the present study (Fig. 2) shows similarities to findings based on the assessment of EB dye and $^3$H-sucrose extravasation in models of focal cerebral ischemia in the rat (Belayev *et al.* 1996;Rosenberg *et al.* 1998;Huang *et al.* 1999). The mechanism of the delayed maximal opening at 48 h remains poorly understood. This second opening is associated with severe ischemic injury, edema and leukocyte infiltration (Figs. 2 and 3). There is also considerable evidence supporting a detrimental role of the delayed neutrophil infiltration to the development of ischemic brain damage (Hartl *et al.* 1996;Matsuo *et al.* 2001;Martin *et al.* 2006). Moreover, tissue swelling ensues within the rigid confines of the skull, elevating intracranial pressure, and ultimately leading to brain herniation and death (Hacke *et al.* 1996). Vascular endothelial leakiness has been proposed to result from the release of cytokines, free radicals, matrix metalloproteinases (MMPs), nitric oxide, histamine, endothelin-1, and products of arachidonic acid metabolism (Wahl *et al.* 1988;Rosenberg *et al.* 1996;Rosenberg *et al.* 1998;Rosenberg 1999;Abbott 2000;Asahi *et al.* 2001;Matsuo *et al.* 2001;Heo *et al.* 2005).

One caveat of the present study is that we didn't elucidate the exact molecular mechanisms through which selective inhibition of COX-2 by nimesulide is able to protect the BBB during reperfusion injury. Elucidation of these mechanisms could explain, in part, the neuroprotective efficacy of COX-2 inhibitors in animal models of stroke, as demonstrated by several research groups. However, since this is the first report to document the ability of a COX-2 inhibitor to protect against ischemia-induced BBB disruption, leukocyte infiltration and edema, it will certainly fuel new investigations aimed at unraveling the mechanism of protection of this new class of COX inhibitors in the context of ischemic stroke.

During the analysis of the data from the present study, and confronting these findings with the scientific literature, several new hypotheses and/or possible mechanisms arose in order to give a plausible explanation to our present findings: *1)* COX-2 inhibition proved to prevent $PGE_2$ formation in the ischemic cortex (Fig. 5), which might be linked to BBB injury. In fact, $PGE_2$ has been previously shown to increase permeability in bovine brain microvessel endothelial cells (BBMEC), which is an *in vitro* model of BBB (Mark *et al.* 2001). In the same study, it was demonstrated that increases in the expression of COX-2 and the release of $PGE_2$ induced by TNF-$\alpha$ were correlated with the permeability and cytoskeletal changes observed in BBMEC in the presence of TNF-$\alpha$. More importantly, it was also shown that inhibition of COX-2 with NS-398 potently reduced TNF-$\alpha$-induced permeability (Mark *et al.* 2001). In support of this study, there is a very recent report indicating that the COX inhibitor ibuprofen completely preserved BBB permeability in an *in* *vitro* BBB model using rat brain microvascular endothelial cells (Krizanac-Bengez *et al.* 2006); *2)* It has long been known that increased production of ROS is related to ischemic microcirculatory injury (Heo *et al.* 2005), and COX-2 activity is a major source of ROS during neuroinflammation both *in vitro* and *in vivo*, as previously demonstrated by our and other groups (Tyurin *et al.* 2000;Pepicelli *et al.* 2002;Akundi *et al.* 2005;Pepicelli *et al.* 2005;Candelario-Jalil *et al.* 2006;Im *et al.* 2006); *3)* COX expression/activity has been implicated in the regulation of endothelial-leukocyte interactions during ischemia at the site of the BBB. In an elegant study by (Stanimirovic *et al.* 1997), it was demonstrated that COX inhibition by indomethacin is able to reduce neutrophil adhesion to human cerebrovascular endothelial cells (HCEC) mediated by several stimuli including exposure to IL-1β and ischemia-like conditions. In the same report, indomethacin completely inhibited IL-1β- and ischemia-induced expression of ICAM-1 by HCEC (Stanimirovic *et al.* 1997); and *4)* A potential link between COX-2 expression / $PGE_2$ formation, and the expression/activity of matrix metalloproteinases, which are involved in BBB damage, should be also considered. This notion is based on recent evidences indicating a $PGE_2$-mediated mechanism involved in MMPs expression by several cell types under inflammatory conditions (Khan *et al.* 2004;Cipollone *et al.* 2005;Pavlovic *et al.* 2006).

In summary, the present study sheds additional light on the neuroprotective effects of the COX-2 inhibitor nimesulide against ischemia-induced $PGE_2$ formation, BBB damage, leukocyte infiltration, and vasogenic edema in a rat model of transient focal cerebral ischemia. It is important to note that this neuroprotective effect of nimesulide was demonstrated with *postischemic* treatment. Furthermore, this study also indicates that COX-1 inhibition is unable to confer any protective effect in focal ischemic brain damage, demonstrating the major contribution of COX-2, rather than COX-1, to brain injury in this model of ischemic stroke. Inhibition of COX-2 may be a valuable therapeutic strategy targeted specifically to the delayed progression of the lesion that occurs in the postischemic phase.

**Acknowledgements:** The authors are grateful to Dr. Mayra Levi (Gautier-Bagó Laboratories) for kindly providing nimesulide for these studies. We would like to thank Noël H. Mhadu and Ms. María de los Angeles Bécquer for expert technical assistance. ECJ was supported by a research fellowship from the Alexander von Humboldt Foundation (Bonn, Germany).

**Table 1.** Effect of the COX-2 inhibitor nimesulide and the COX-1 inhibitor Valeryl Salicylate (VAS) on total, cortical and subcortical infarct volumes evaluated after 3 days of reperfusion following 1 h of MCAO in the rat.

| Treatment | Infarct Volume ($mm^3$) | | |
|---|---|---|---|
| | **Total** | **Cortical** | **Subcortical** |
| Vehicle (n=7) | $291.6 \pm 53.4$ | $242.5 \pm 49.8$ | $50.4 \pm 9.6$ |
| Nimesulide 12 mg/kg; administered immediately after stroke (n=5) | $160.6 \pm 8.3$ * | $125.6 \pm 7.7$ * | $33.8 \pm 13.1$ * |
| Nimesulide 12 mg/kg; 6 h delayed treatment (n=8) | $150.2 \pm 21.4$ * | $111.5 \pm 26.2$ * | $42.2 \pm 18.8$ |
| Valeryl Salicylate 120 mg/kg; administered immediately after stroke (n=6) | $279.7 \pm 70.6$ | $236.8 \pm 59.9$ | $48.1 \pm 18.7$ |
| Valeryl Salicylate 120 mg/kg; 6 h delayed treatment (n=6) | $296.9 \pm 46.9$ | $239.1 \pm 39.2$ | $57.6 \pm 12.8$ |

*$p<0.05$ with respect to vehicle treated animals. Statistical analysis was performed using ANOVA followed by the Student-Newman-Keuls *post-hoc* test (multiple comparisons) or *t*-test (for detecting individual differences between two groups).

**Table 2.** Effects of COX inhibitors on stroke-induced leukocyte infiltration and edema. The COX-2 inhibitor nimesulide significantly reduced cortical leukocyte infiltration seen after 48 h of reperfusion (assessed by the MPO activity), and potently diminished edema formation in the ischemic brain.

| Treatment | Cortical MPO activity (U/g tissue) | Subcortical MPO activity (U/g tissue) | Edema (%) |
|---|---|---|---|
| Sham-operated (n=5) | $0.10 \pm 0.15$ | $0.11 \pm 0.23$ | - |
| Vehicle (n=8) | $2.95 \pm 0.89$ | $1.94 \pm 0.92$ | $7.62 \pm 1.58$ |
| Nimesulide 12 mg/kg; administered immediately after stroke (n=8) | $1.46 \pm 0.86$ ** | $1.92 \pm 0.77$ | $1.75 \pm 0.89$ ** |
| Nimesulide 12 mg/kg; 6 h delayed treatment (n=8) | $1.68 \pm 0.87$ * | $1.82 \pm 0.84$ | $2.77 \pm 1.61$ * |
| Valeryl Salicylate 120 mg/kg; administered immediately after stroke (n=6) | $3.07 \pm 0.92$ | $1.78 \pm 0.84$ | $6.50 \pm 2.25$ |
| Valeryl Salicylate 120 mg/kg; 6 h delayed treatment (n=7) | $2.75 \pm 1.08$ | $1.90 \pm 1.27$ | $5.94 \pm 1.74$ |

*p<0.05 and **p<0.01 with respect to vehicle-treated animals. ANOVA followed by the Student-Newman-Keuls *post-hoc* test (multiple comparisons) or *t*-test (for detecting individual differences between two groups).

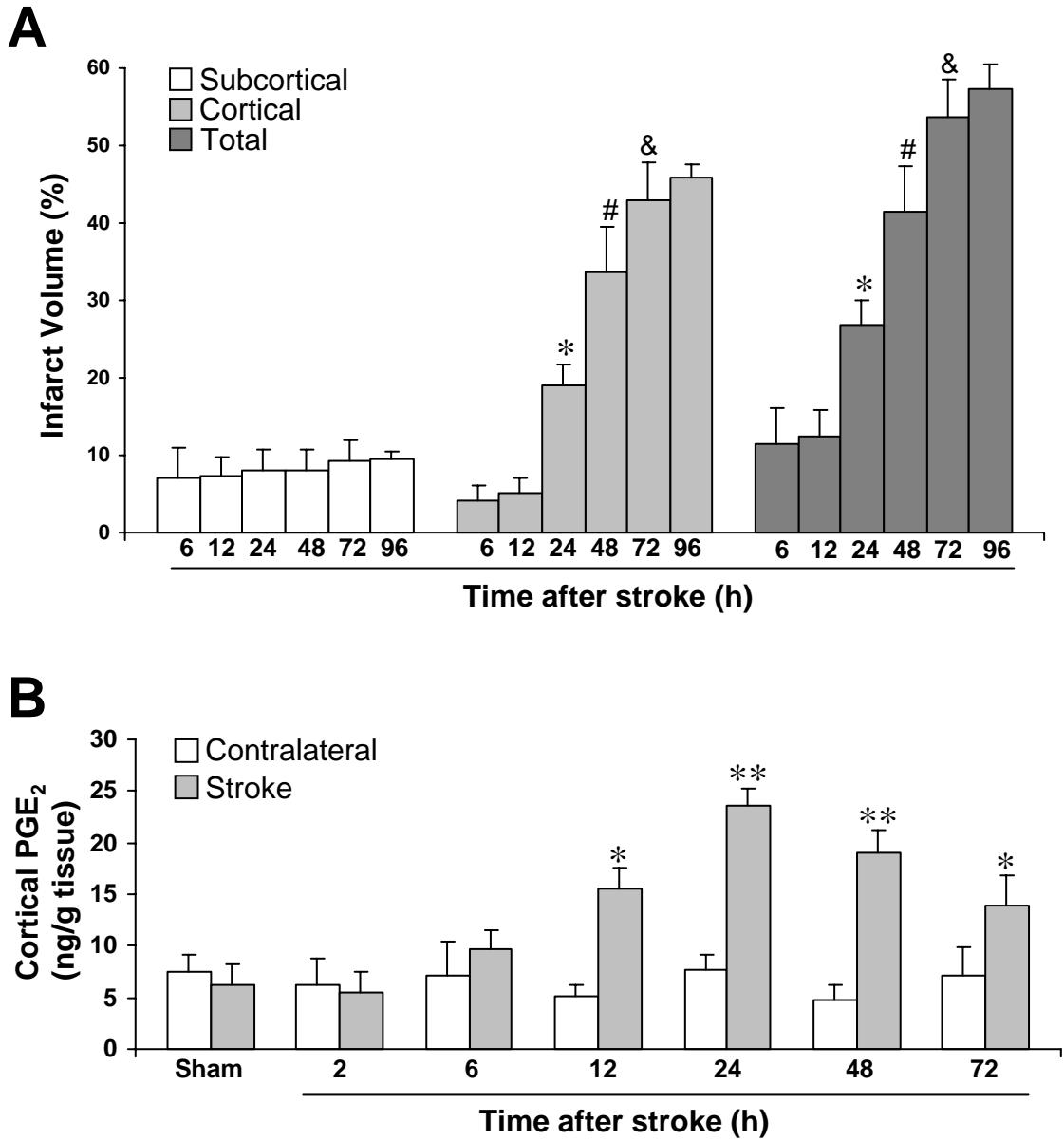

Fig. 1. Temporal evolution of the ischemic lesion (A), and PGE2 production (B) in the rat brain (1 h MCAO and different times of recirculation). Infarct volumes were calculated from six coronal TTC-stained brain slices, and assessed in the cerebral cortex and subcortical regions. Brain damage progresses several hours/days, and is completed by 3 days of recirculation. There is also a delayed production of PGE2 in the ischemic cortex, reaching maximal values by 24 h of reperfusion. In Panel A, *p<0.05 with respect to 12 h; #p<0.05 with respect to infarct volume at 24 h; &p<0.05 with respect to the lesion size at 48 h. In Panel B, *p<0.05 and **p<0.01 with respect to contralateral at a given time point. ANOVA followed by the Student-Newman-Keuls *post-hoc* test (multiple comparisons) or *t*-test (for detecting individual differences between two groups). N=5-9 per time point.

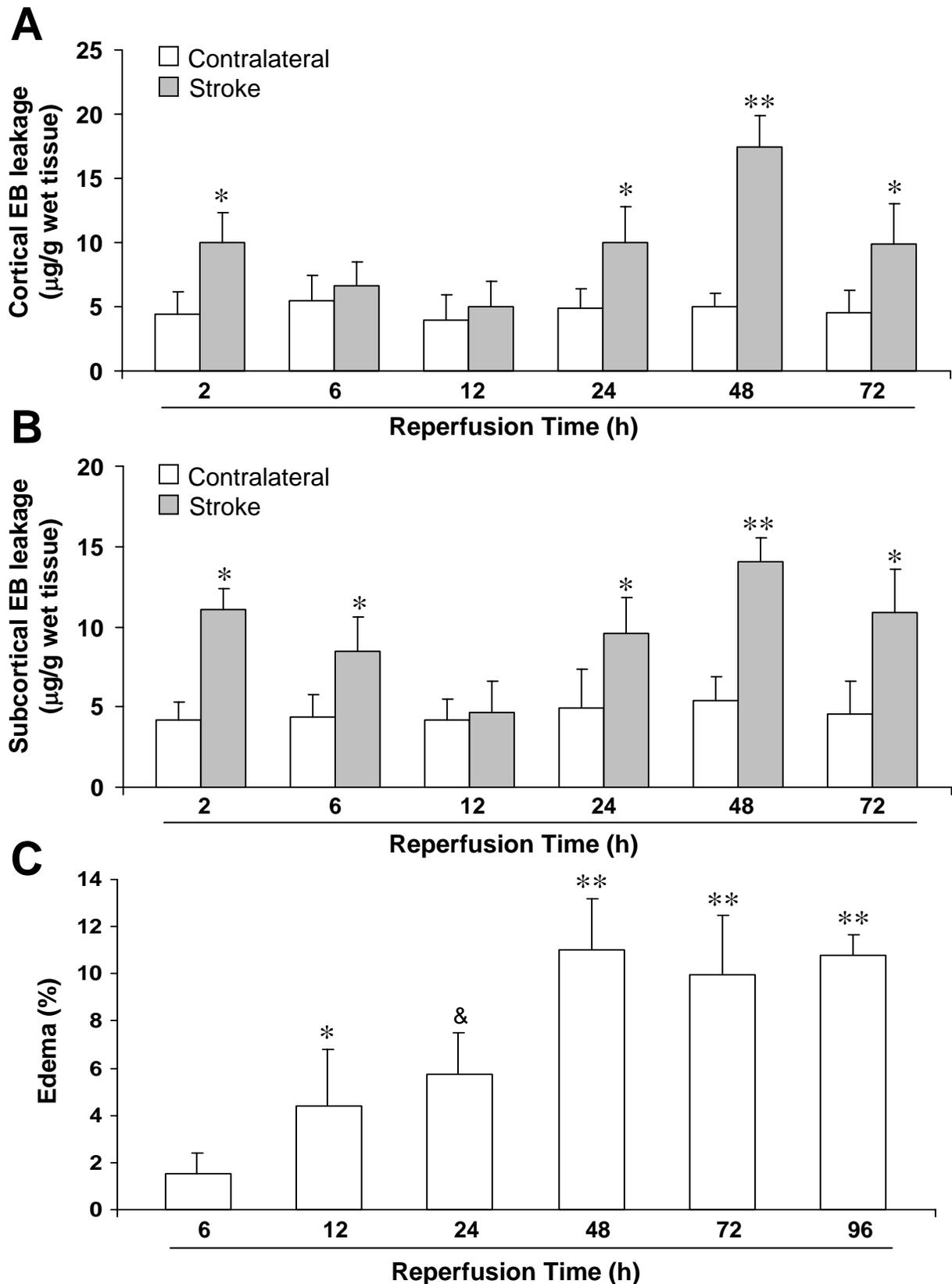

**Fig. 2.** Evaluation of BBB disruption (**A** and **B**), and edema formation (**C**) at different times of reperfusion following MCAO in the rat. BBB breakdown was assessed by quantifying the concentration of Evans Blue leakage into the cerebral cortex (**A**) and subcortical areas (**B**). Edema index was calculated by dividing the total volume of the hemisphere ipsilateral to MCAO by the total volume of the contralateral hemisphere (Yang et al., 1998). In Panel A and B, *$p<0.05$ and **$p<0.01$ with respect to the contralateral side at a particular time point. In Panel C, *$p<0.05$ with respect to 6 h; &$p<0.05$ with respect to 12 h; **$p<0.01$ with respect to 24 h. ANOVA followed by the Student-Newman-Keuls *post-hoc* test (multiple comparisons) or *t*-test (for detecting individual differences between two groups). N=5-9 per time point.

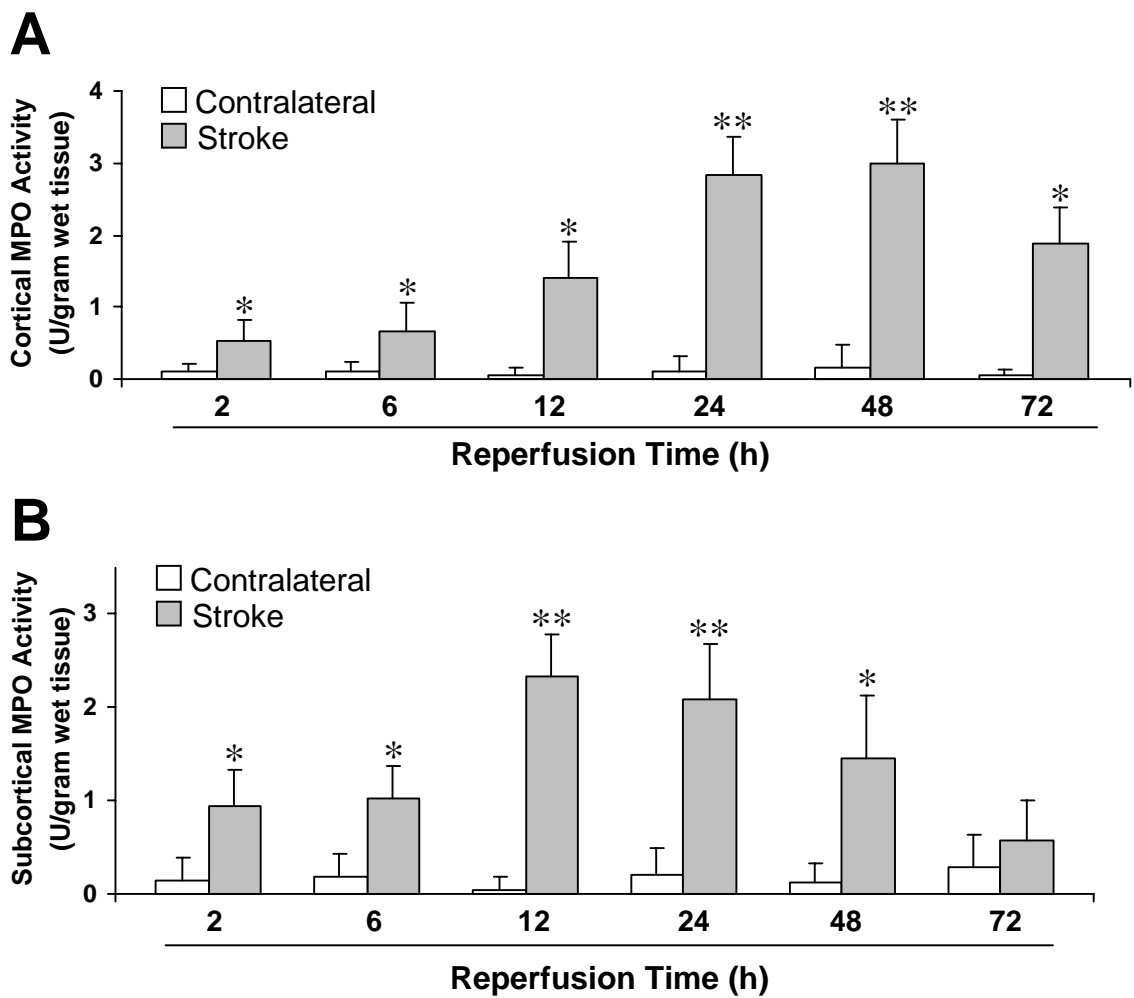

**Fig. 3.** Time course of leukocyte infiltration into the ischemic brain. Myeloperoxidase (MPO) activity was evaluated in the cortical areas (A) and in the subcortex (B) in ischemic and contralateral sides at different times after removal of the filament occluding the MCA in the rat. *p<0.01 and **p<0.001 with respect to the contralateral MPO activity. Statistical analysis was performed using ANOVA followed by the Student-Newman-Keuls *post-hoc* test (multiple comparisons) or *t*-test (for detecting individual differences between two groups). N=5-9 per time point.

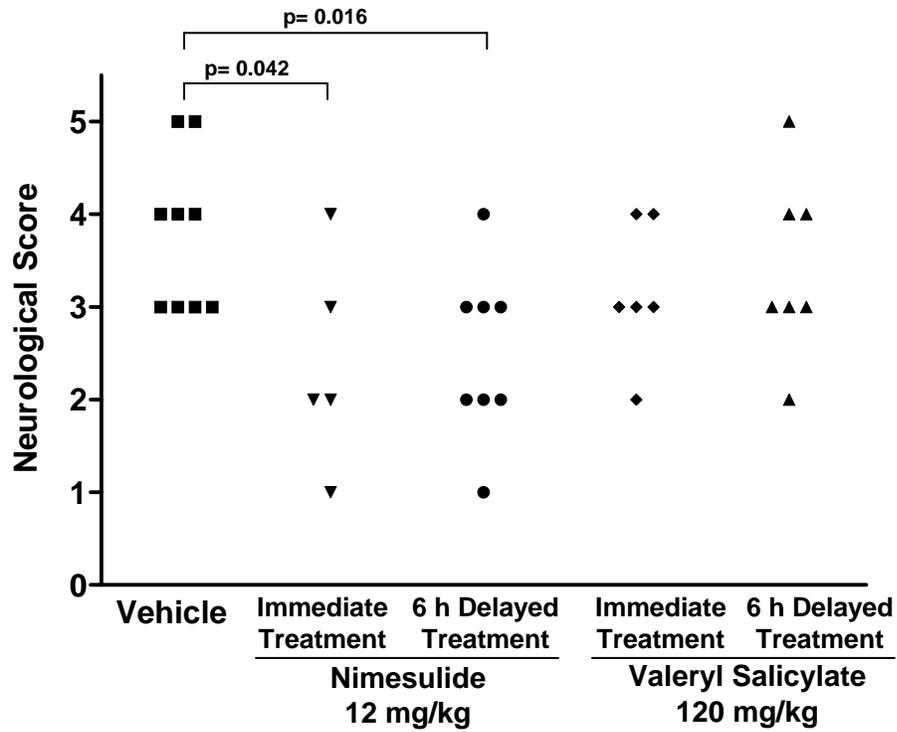

**Fig. 4.** Scatter plots of neurological deficit scores in each treatment group evaluated at 3 days after the induction of transient focal cerebral ischemia in the rat. Statistical analysis was performed using the Mann-Whitney nonparametric test. N=5-9 per treatment group.

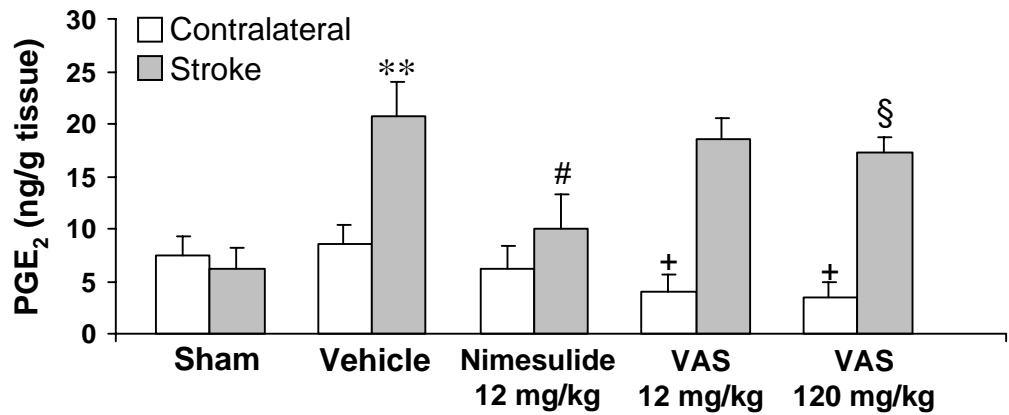

**Fig. 5**. Production of PGE2 in the ischemic cerebral cortex is potently reduced by the COX-2 inhibitor nimesulide, but only very modestly diminished by the highest dose of the COX-1 inhibitor valeryl salicylate (VAS). PGE2 levels were determined using an enzyme immunoassay after 24 h of reperfusion following 1 h of ischemia. **$p < 0.01$ with respect to the contralateral side; #$p < 0.01$ and §$p < 0.05$ with respect to the stroke side of vehicle-treated animals; +$p < 0.05$ with respect to sham-operated rats. ANOVA followed by the Student-Newman-Keuls *post-hoc* test (multiple comparisons) or *t*-test (for detecting individual differences between two groups). N=5-7 animals per group.

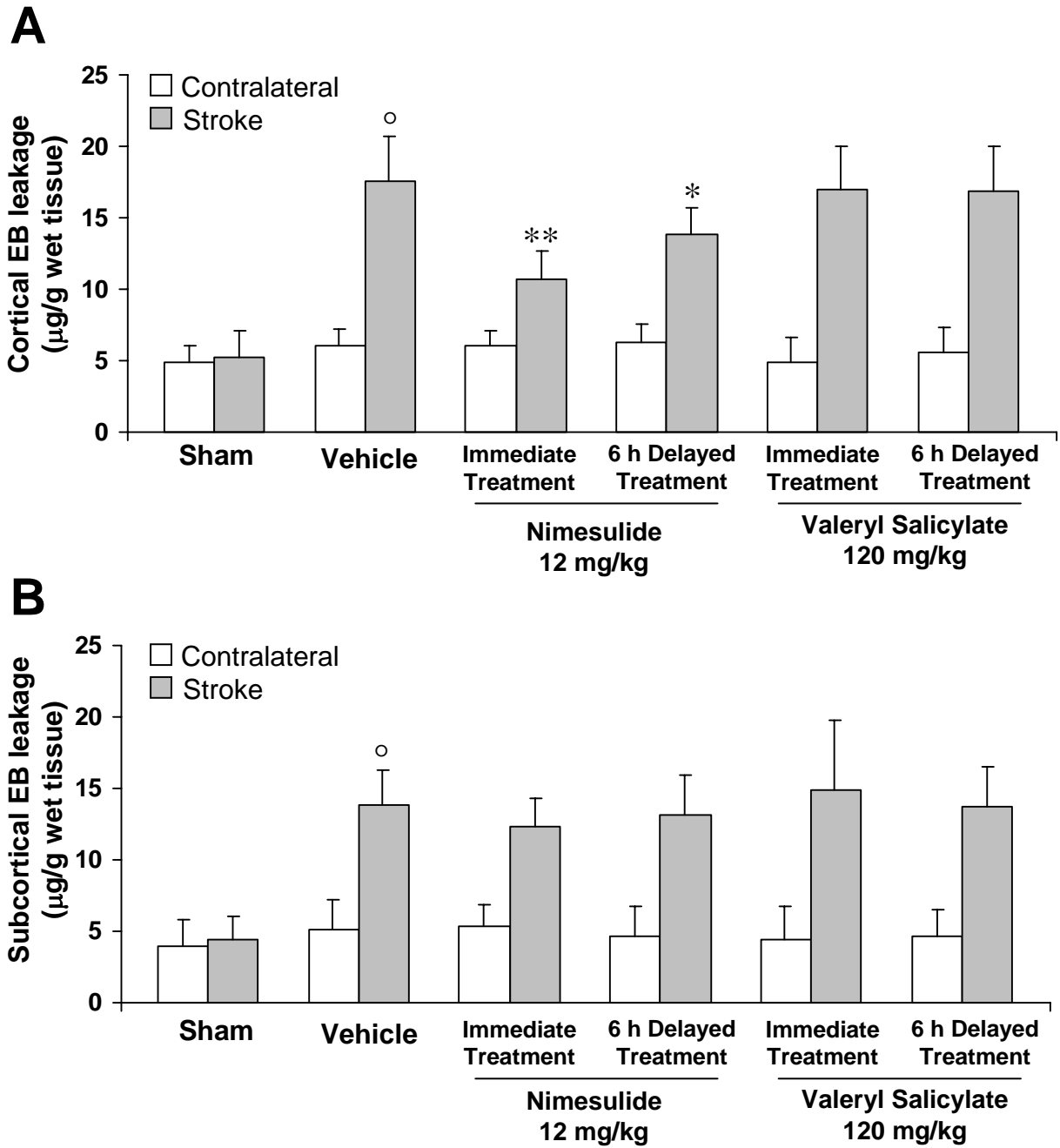

**Fig. 6.** Effects of nimesulide and VAS on the damage to the BBB, as assessed by the Evans blue (EB) extravasation method. Concentrations of EB were determined in the ischemic cerebral cortex (A) and subcortical areas (B). °p<0.001 with respect to the contralateral side; **p<0.01 and *p<0.05 with respect to the stroke side of vehicle-administered animals. Determination of statistical differences among treatment groups was performed using ANOVA followed by the Student-Newman-Keuls *post-hoc* test (multiple comparisons) or *t*-test (for detecting individual differences between two groups). N=6-12 animals per group.